\documentclass[a4paper,10pt]{revtex4}
\usepackage{mathrsfs}
\usepackage{graphicx}
\usepackage{latexsym}
\usepackage{amsmath}
\usepackage{amssymb}
\usepackage{textcomp}
\usepackage{amsbsy}
\usepackage{graphics}
\usepackage{epstopdf}
\usepackage{color}

\allowdisplaybreaks[4]

\begin{document}

\tolerance=5000

\title{The effect of thermal radiation on singularities in the dark universe}

\author{I.~Brevik,$^{1}$\,\thanks{iver.h.brevik@ntnu.no}
A.~V.~Timoshkin,$^{2,3}$\,\thanks{alex.timosh@rambler.ru}
Tanmoy~Paul$^{4,5}$\,\thanks{pul.tnmy9@gmail.com}} \affiliation{ $^{1)}$ Department of Energy and Process Engineering,
Norwegian University of Science and Technology, N-7491 Trondheim, Norway\\
$^{2)}$ Tomsk State Pedagogical University, Kievskaja Street, 60, 634061 Tomsk, Russia \\
$^{3)}$ International Laboratory of Theoretical Cosmology, Tomsk State University of Control Systems and Radio Electronics,
Lenin Avenue, 36, 634050 Tomsk, Russia\\
$^{4)}$ Department of Physics, Chandernagore College, Hooghly - 712 136.\\
$^{(5)}$ Department of Theoretical Physics, Indian Association for the Cultivation of Science, 2A $\&$ 2B Raja S.C. Mullick Road, Kolkata - 700 032, India }

\tolerance=5000

\begin{abstract}
Cosmological models with an inhomogeneous viscous dark fluid, coupled with dark matter in the Friedmann-
Robertson-Walker (FRW) flat universe, are considered. The influence of thermal effects caused by Hawking radiation
on the visible horizon is studied, in connection with the classified type I and type III singularities which are known to occur within a finite amount of time. Allowance of thermal effects implies that a transition to a type II singularity can take place, in a finite time. We take into account a bulk viscosity of the dark fluid, observing the equation of state in the case of radiation, and find that there is a qualitative change in the singular universe of type I: it may pass into a singularity of type III, or it may avoid the singularity at all.

\end{abstract}

%\pacs{}

\maketitle

\section{Introduction}
After it was observed  that the universe is exposed to an accelerated expansion,  it is of great interest
to study the nature of dark energy which is responsible for this acceleration
\cite{Nojiri:2017ncd,Shi:2011sa,Perivolaropoulos:2006ce,Arefeva:2006ido}.
In the era of dark energy  the universe can be qualitatively described using an exotic effective
fluid with negative pressure (more appropriately called a positive tensile stress), that satisfies an unusual equation of state \cite{Nojiri:2005sx,Nojiri:2006zh,
Nojiri:2004pf,Setare:2007bx,Capozziello:2005mj,Cognola:2006eg,Nojiri:2005sr,Nojiri:2020wmh,Odintsov:2020zct}. Dark energy
can be characterized by an equation of state parameter $\omega_\mathrm{eff} = p_\mathrm{eff}/\rho_\mathrm{eff}$ . Depending on the values of $\omega_\mathrm{eff}$ several possibilities of
the universe are possible: for example, phantom behaviour occurs for $\omega_\mathrm{eff} < -1$, de-Sitter evolution occurs for $\omega_\mathrm{eff} = -1$,
and quintessential behaviour occurs for $-1 < \omega_\mathrm{eff} < -1/3$. The
experimental value of the parameter of the equation of state is however determined with insufficient accuracy
to be able to unambiguously determine the phase in which our Universe is located. Today
the value of this parameter lies within the following limits $\omega_\mathrm{eff} = 1.04^{+0.09}_{-0.10}$ \cite{Nakamura:2010zzi}.

One of the properties of phantom dark energy is the prediction of a  Big Rip singularity (type I)
in the future.  For this kind of singularity the scale factor and the Hubble function go to infinity at a finite time, called $t_s$ or
 $t_\mathrm{rip}$. This is the most destructive type of singularity \cite{Caldwell:1999ew}. There are several less
drastic  future singularities, namely soft singularities classified  as   type II, III and IV. All these singularities are purely classic in  nature.

From a physical viewpoint, an increase in the Hubble function should be expected to lead to  an increase in  temperature.
 At high temperatures, especially near the singularity, thermal radiation should appear.
Thermal radiation is associated with Hawking radiation, which effectively should be generated at
the apparent horizon of the FRW universe \cite{Cai:2009ph}. Hawking's radiation is manifested in black
holes and is associated with the existence of the visible black hole horizon, as well as the visible
horizon of cosmic events in de-Sitter space. Hawking's thermal spectrum radiation should be taken to appear
in the late universe at high temperatures shortly before its rupture. Accounting for thermal radiation
will allow a qualitative change in the classical description and give a more realistic picture of the
future of the universe.

Recently, the effect of thermal radiation on future singularities of types  I, II, III and IV  was
 studied in Ref.~\cite{Nojiri:2020sti}. It was shown that with singular universes of types I and III, as well as for
the Little Rip universe,   there occurs  a qualitative change in the  singularity due to thermal effects.
The singularities  end up as type II singularities. In  universes of types II and IV  there is no qualitative
change in the final state.

The model of a non-viscous fluid in cosmology is an idealized case.
Dark energy universe with a viscous  fluid was studied in Refs. \cite{Gron:1990ew,Brevik:2017msy,Cataldo:2005qh,Brevik:2005gx,
Brevik:2001ed,Brevik:2010jv,Brevik:2005bj,Li:2009mf,Brevik:2017juz}. The cosmic
viscosity property was taken into account in connection with  the Big Rip type singularity \cite{Nojiri:2003vn}, and also
 in connection with the singularities of type II, III and IV  \cite{Barrow:2013ria,Nojiri:2004ip}. Models of dark energy interacting with dark
matter, in which the singularities of one of these four types are formed, were considered in Ref.~\cite{Myrzakul:2013qka,Capozziello:2009hc}.

The article \cite{Elizalde:2014ova} studied a combined phantom/fluid model consisting of a viscous
dark fluid (dark energy) with a linear inhomogeneous equation of state and dark matter with a
linear homogeneous equation of state in a spatially flat FRW metric.

The purpose of the present article is to study the influence of thermal radiation, taking into
account the interaction of viscous dark fluid with dark matter. In particular, we will focus on  the change of singular behavior
of the late-time universe.

\section{The effect of thermal radiation on the formation of singularities in the late universe}\label{sec_2}

Let us consider a spatially-flat FRW universe
\begin{eqnarray}
 ds^2 = -dt^2 + a^2(t)\delta_{ij}dx^{i}dx^{j},
 \label{1}
\end{eqnarray}
where $a(t)$ is the scale factor. We will be interested in the case when the effective parameter in
the equation of state takes values in the vicinity of $-1$. Then the following kinds of evolution of
the accelerating universe are possible : phantom, quintessence, or a de Sitter expansion. The question
arises about  how  this evolution will end up in the future.  The answer depends on the behavior of the
time-dependent parameters of the effective equation of state. We will be interested
in dark energy universes, in which there occur   future singularities within a finite, or an infinite,  time. For such universes the Nojiri-Odintsov-Tsujikawa classification
was given in Ref.~\cite{Nojiri:2005sx} (see also \cite{Odintsov:2018uaw}). Singularities arise when one or more of the central cosmological parameters diverge:  the scale factor $a(t)$, the effective (total) energy density
$\rho_\mathrm{eff}$, the effective (total) pressure $p_\mathrm{eff}$, or higher derivatives of the Hubble function.

In the limit $t \rightarrow t_s$, the following classes of singularities can be distinguished:

\begin{itemize}
 \item Type I (Big Rip): $a \rightarrow \infty$, $\rho_\mathrm{eff}\rightarrow \infty$ and $p_\mathrm{eff} \rightarrow \infty$.
 This class of singularities includes the case when $\rho_\mathrm{eff}$ and $p_\mathrm{eff}$ are finite at $t = t_s$.
 A Big Rip leads to the decay of gravitationally bound objects large on a cosmological scale.

 \item Type II (“sudden” singularity): $a \rightarrow a_s$ , $\rho_\mathrm{eff} \rightarrow \rho_s$ and $\big|p_\mathrm{eff}\big| \rightarrow \infty$,
 where $a_s \neq 0$ and $\rho_s$ are constant. That is a pressure singularity.

 \item Type III: $a \rightarrow a_s$, $\rho_\mathrm{eff} \rightarrow \infty$ and $p_\mathrm{eff} \rightarrow \infty$.
 This type singularity is milder than Type I but stronger than Type II.

 \item Type IV: $a \rightarrow a_s$, $\rho_\mathrm{eff} \rightarrow 0$ and $p_\mathrm{eff} \rightarrow 0$,
 but the higher derivatives of the Hubble function $H$ diverge. This type also includes the case where $\rho_\mathrm{eff}$ and/or $p_\mathrm{eff}$
 are finite for $t = t_s$.
\end{itemize}

Here $\rho_\mathrm{eff}$ and $p_\mathrm{eff}$ can be calculated by the following expressions,
\begin{eqnarray}
 \rho_\mathrm{eff} = \frac{3}{\kappa^2}H^2~~~~~~~~~~~,~~~~~~~~~~~p_\mathrm{eff} = -\frac{2}{\kappa^2}\big(2\dot{H} + 3H^2\big)
 \label{2}
\end{eqnarray}
where $\kappa^2= 8\pi G$ and $H = \frac{\dot{a}}{a}$ is the Hubble parameter. It may be mentioned that the effective energy density $\rho_\mathrm{eff}$ and
the effective pressure $p_\mathrm{eff}$ may include the contribution from the modified gravity.

However, the singularity is not the only possible outcome of the evolution of our universe
in the phantom phase. It was shown in \cite{Odintsov:2018uaw,Frampton:2011aa,Frampton:2011sp,Frampton:2011rh} that if the cosmic energy density
remains constant or monotonically increases, then, depending on the asymptotic behavior of the
Hubble parameter $H$ \cite{Odintsov:2018uaw}, all possible types of evolution of our universe can be divided into four categories:

\begin{itemize}
 \item Big Rip: $H(t) \rightarrow \infty$ when $t = t_s < \infty$.

 \item Little Rip: $H(t) \rightarrow \infty$ when $t \rightarrow \infty$.

 \item Cosmological constant : $H(t) = \mathrm{constant}$.

 \item Pseudo Rip: $H(t) \rightarrow H_{\mathrm{}\infty}$ when $t \rightarrow \infty$, where $H_\mathrm{\infty}$ is a constant.
\end{itemize}

Here we would like to mention that both the Little Rip and Pseudo Rip models are nonsingular.

We will study the cosmological models induced by the inhomogeneous viscous dark fluids
coupled with dark matter, in terms of the parameters appearing in the equation of state (EoS).
Let us consider the following formulation of the EoS of an inhomogeneous viscous fluid in flat FRW space-time \cite{Capozziello:2005pa}, namely
\begin{eqnarray}
 p = \omega(\rho,t)\rho - 3H\xi(H,t)~~~,
 \label{3}
\end{eqnarray}
where $\xi(H,t)$ is the bulk viscosity, which depends on the Hubble parameter $H$ and on the cosmic time $t$.
According to the thermodynamic set up, we naturally assume that $\xi(H,t) > 0$.

We will take the following form for the thermodynamic (EoS) parameter $\omega$ \cite{Capozziello:2005pa},
\begin{eqnarray}
 \omega(\rho,t) = \omega_1(t)\big(A_0\rho^{\alpha - 1} - 1\big)
 \label{4}
\end{eqnarray}
where $A_0 \neq 0$ and $\alpha \geq 1$ are constants. (Note that $A_0$ is nondimensional only if $\alpha= 1$. If $\alpha= 3/2$ for instance, the dimension will be cm$^2$ in geometric units.)

 We choose the bulk viscosity as \cite{Capozziello:2005pa},
\begin{eqnarray}
 \xi(H,t) = \xi_1(t)\big(3H\big)^n
 \label{5}
\end{eqnarray}
with $n > 0$.

Let us consider the influence of thermal effects on the change in singularities of types I,
taking into account the viscosity property of a dark fluid and its interaction with dark matter.
Since the temperature of the universe increases near the singularity, thermal radiation is generated, as mentioned above.
From  statistical physics, the energy density of thermal radiation is
proportional to the fourth power of the absolute temperature. Therefore, near the future singularity where
the Hubble parameter becomes very high, we assume  that the thermal energy density  has the form \cite{Nojiri:2020sti}
\begin{eqnarray}
 \rho_\mathrm{rad} = \lambda H^4~~,
 \label{6}
\end{eqnarray}
where $\lambda$ is a positive constant.

Taking thermal radiation into account, the FRW equation is modified as follows \cite{Nojiri:2020sti},
\begin{eqnarray}
 \frac{3}{\kappa^2}H^2 = \rho_\mathrm{eff} + \lambda H^4
 \label{7}
\end{eqnarray}
From Eq.(\ref{7}) it follows that when the evolution time of the late universe is much less than
the singularity time, the first term of the equation makes the greatest contribution. While near the
singularity time, the second term makes the largest contribution. We analyze equation (7) further, by solving it  with respect to the square of the Hubble parameter $H^2$
\begin{eqnarray}
 H^2 = \frac{1}{2\lambda}\bigg[\frac{3}{\kappa^2} \pm \sqrt{\frac{9}{\kappa^4} - 4\lambda\rho_\mathrm{eff}}\bigg]~~.
 \label{8}
\end{eqnarray}
In the following, we will apply the cosmological models of a viscous fluid from the article \cite{Elizalde:2014ova}.

\section{Singular behavior of late-time Universe taking into account the viscosity of a fluid and its interaction with dark matter}\label{sec_3}
We start from the simplest, constant case, namely $\omega(\rho,t) = \omega_0$ and will consider different forms for the bulk viscosity.

\subsection{Constant viscosity}
Let us consider the case of constant bulk viscosity $\xi(H,t) = \xi_0 > 0$. The Hubble function
 has the form \cite{Elizalde:2014ova}

\begin{eqnarray}
 H(t) = \frac{\xi_0\kappa^2}{\big(1 + \omega_0\big)\bigg(1 - \sqrt{C_1}~\exp{\big[3\xi_0\kappa^2t/2\big]}\bigg)},
 \label{9}
\end{eqnarray}
with $C_1$ a nondimensional constant.
This model does not so far take into account interactions with dark matter. Observe that $H$ diverges for
$t \rightarrow t_s = -\frac{2}{3\xi_0\kappa^2}\ln{\big(\sqrt{C_1}\big)}$, thus a
Big Rip singularity appears. Let us  now see if the type of singularity will change if we take into account thermal radiation near the singularity time.

We calculate the scale factor,
\begin{eqnarray}
 a(t) = e^{\int Hdt} = a_0\bigg(1 - \frac{1}{\sqrt{C_1}}~\exp{\big[-3\xi_0\kappa^2t/2\big]}\bigg)^{\frac{2}{3(1+ \omega_0)}}~~,
 \label{10}
\end{eqnarray}
where $a_0$ is an integration constant. Then the effective energy density in terms of the scale factor is
\begin{eqnarray}
 \rho_\mathrm{eff} = \bigg(\sqrt{\frac{3}{\rho_0}}~\frac{\xi_0\kappa}{1+\omega_0}\bigg)^2\bigg[\frac{a(t)}{a_0}\bigg]^{3(1+\omega_0)}~e^{-3\xi_0\kappa^2t}~~.
 \label{11}
\end{eqnarray}
Let us  return to Eq.(\ref{8}). Since $H^2$ is a real number, we must have
\begin{eqnarray}
 \frac{9}{\kappa^4} - 4\lambda\rho_\mathrm{eff} \geq 0~~.
 \label{12}
\end{eqnarray}
We get in this model
\begin{eqnarray}
 \frac{9}{\kappa^4} - 4\lambda A e^{-Bt}\bigg[\frac{a(t)}{a_0}\bigg]^{C} \geq 0~~,
 \label{13}
\end{eqnarray}
where $A = \bigg(\sqrt{\frac{3}{\rho_0}}\frac{\xi_0\kappa}{1+\omega_0}\bigg)^2$, $B = 3\xi_0\kappa^2$ and $C = 3(1 + \omega_0)$. The above inequality
puts a restriction on the scale factor,
\begin{eqnarray}
 a(t) \leq a_0\bigg(\frac{9}{4\lambda A\kappa^4}\bigg)^{\frac{1}{C}}~e^{\frac{B}{C}t}
 \label{14}
\end{eqnarray}
under the condition $C < 0$ ($\omega_0 < -1$), which corresponds to the case of phantom dark energy.

Taking into account the thermal radiation $\rho_{\rm rad}$, we obtain from Eq.~(\ref{8}) that there exists another upper limit $a_{\rm max}$ for the scale parameter,
\begin{eqnarray}
 a(t) \leq a_\mathrm{max} = a_0\bigg(\frac{4\lambda A\kappa^4}{9}\bigg)^{\frac{1}{3(1+\omega_0)}}~~,
 \label{15}
\end{eqnarray}
which corresponds to the instant $t_\mathrm{max}$,
\begin{eqnarray}
 t_\mathrm{max} = -\frac{2}{3\xi_0\kappa^2}\ln{\bigg\{\sqrt{C_1}\bigg(1 - \frac{2\kappa^2\sqrt{A\lambda}}{3}\bigg)\bigg\}}~~.
 \label{16}
\end{eqnarray}
This is thus a singularity of another type than that arising from the Hawking radiation.

We calculate the difference between $t_\mathrm{max}$ and $t_s$
\begin{eqnarray}
 t_\mathrm{max} - t_s = -\frac{2}{3\xi_0\kappa^2}\ln{\bigg\{1 - \frac{2\kappa^2\sqrt{A\lambda}}{3}\bigg\}} > 0~~,
 \label{17}
\end{eqnarray}
which shows  that $t_\mathrm{max}$ is larger than $t_s$. In the limit $t \rightarrow t_\mathrm{max}$, $a \rightarrow a_\mathrm{max}$.
From Eqs.(\ref{2}) and (\ref{3}), one can calculate the effective energy
density $\rho_\mathrm{eff}$ and the effective pressure $p_\mathrm{eff}$. Then in the limit $t \rightarrow t_\mathrm{max}$, the effective energy density
and effective pressure become
\begin{eqnarray}
 \rho_\mathrm{max} = \rho_\mathrm{eff}(t_\mathrm{max}) = \frac{9\xi_0^2\kappa^2}{(1+\omega_0)^2}\bigg[\frac{1}
 {1 + C_1\bigg(\frac{2\kappa^2\sqrt{A\lambda}}{3} - 1\bigg)}\bigg]^2~~,
 \label{18}
\end{eqnarray}
and
\begin{eqnarray}
 p_\mathrm{max} = p_\mathrm{eff}(t_\mathrm{max}) = \bigg|\omega_0\rho_\mathrm{max} - 3\xi_0H_\mathrm{max}\bigg| =
 \frac{3\xi_0^2\kappa^2}{(1+\omega_0)}\bigg\{\frac
 {\bigg|\frac{2\omega_0 - 1}{1+\omega_0} - C_1\bigg(\frac{2\kappa^2\sqrt{A\lambda}}{3} - 1\bigg)\bigg|}
 {\bigg[1 + C_1 \bigg(\frac{2\kappa^2\sqrt{A\lambda}}{3} - 1\bigg)\bigg]^2}\bigg\}~~.
 \label{19}
\end{eqnarray}
respectively. In the general case the values of the scale factor, energy density and effective pressure turn
out to be finite, but higher derivatives of H diverge. Thus, a cosmological  finite-time future
singularity is not formed. This behavior is due to the influence from the viscosity of the dark
fluid, which compensates for the effect of thermal radiation. However, if the radiation parameter $\lambda$ goes to $\lambda_0$, with
\begin{eqnarray}
 \lambda_0 = \frac{3}{C_1}\bigg(\frac{(C_1 - 1)(1 + \omega_0)}{2\xi_0\kappa^3}\bigg)^2~~,
 \label{20}
\end{eqnarray}
then $\rho_\mathrm{max} \rightarrow \infty$ and $\big|p_\mathrm{max}\big| \rightarrow \infty$.
This is a type III singularity. It is milder than type I, but stronger than type II.

If $\zeta_0 \rightarrow \infty$, then $\lambda_0 \rightarrow 0$. Thus, the viscosity weakens the effect of thermal radiation.

\subsection{Viscosity proportional to the Hubble parameter}

Let us consider the case where the viscosity is proportional to  the Hubble parameter,
 $\xi(H,t) = 3\tau H$, the constant $\tau$ being positive. The Hubble parameter becomes \cite{Elizalde:2014ova}
\begin{eqnarray}
 H(t) = \frac{\kappa}{\sqrt{3}}\bigg\{\frac{\delta{\gamma}~\sqrt{C_1}}{3\theta\sqrt{C_1} + \exp{\big[-\tilde{\eta}t/2\big]}}\bigg\}~~.
 \label{21}
\end{eqnarray}
Here the following designations are introduced:
\[ \eta = \delta \gamma^2, \quad \gamma = \frac{\kappa}{\sqrt 3}\sqrt{ 1+\frac{1}{r}}, \quad \frac{\tilde{\eta}}{\eta}=r, \quad \theta = 1 + \omega_0 - 9\tau {\gamma}^2, \]
 where the constant $r$ is associated with the influence of dark matter and is equal to the ratio of the energy density of dark matter to the density of dark energy. The dimensions are $[\gamma]=~$cm, $[\eta] =[\tilde{\eta}]=$ cm$^{-1}$, $[\delta]= $cm$^{-2}$, $[\tau]=~$cm$^{-2}$.
 Further, the   constant parameter $\delta$ is responsible for the interaction with dark matter. If  $\omega_0 < -1 + 9\tau{\gamma}^2$, then $\theta < 0$ and consequently $H$ diverges at
$t \rightarrow t_s = -\frac{2}{\tilde{\eta}}\ln{\bigg(-3\theta\sqrt{C_1}\bigg)}$, which leads to the appearance of a singularity of the type Big Rip.
We consider again the behavior of the late-time universe near the singularity, taking into account the effect of thermal radiation.

Let us calculate the scale factor,
\begin{eqnarray}
 a(t) = a_0~\exp{\big[\frac{\delta{\gamma}~\kappa t}{3\sqrt{3}~\theta}\big]}\bigg(3\theta\sqrt{C_1} + \exp{\big[-\tilde{\eta}t/2\big]}
 \bigg)^{\frac{2\delta{\gamma}~\kappa}{3\sqrt{3}~\theta\tilde{\eta}}}
 \label{22}
\end{eqnarray}
and express the effective energy density in terms of the scale factor as
\begin{eqnarray}
 \rho_\mathrm{eff} = C_1~\big(\delta {\gamma}\big)^2\bigg[\frac{a(t)}{a_0}\bigg]^{-\frac{2}{\tilde{\delta}}}~
 \exp{\big[\frac{2\alpha^\prime}{\tilde{\delta}}t\big]}~~,
 \label{23}
\end{eqnarray}
where $\alpha^\prime = \frac{\delta \gamma \kappa}{3\sqrt{3}\theta}$ and
$\tilde{\delta} = \frac{2\delta\gamma\kappa}{3\sqrt{3}\theta}$. The
inequality in Eq.(\ref{12}) implies the following restriction on the scale factor,
\begin{eqnarray}
 a(t) \leq a_0~\bigg(\frac{2\delta{\gamma}~\kappa^2\sqrt{\lambda C_1}}{3}\bigg)^{\tilde{\delta}}~e^{\alpha^\prime t}~~.
 \label{24}
\end{eqnarray}
Since $\theta < 0$, then $\alpha^\prime < 0$ and scale factor values are limited by the maximum number $a_\mathrm{max}$ given by
\begin{eqnarray}
 a(t) \leq a_\mathrm{max} = a_0~\bigg(\frac{2\delta{\gamma}~\kappa^2\sqrt{\lambda C_1}}{3}\bigg)^{\tilde{\delta}}~~,
 \label{25}
\end{eqnarray}
which corresponds to the instant
\begin{eqnarray}
 t_\mathrm{max} = \frac{2}{\tilde{\eta}}~\ln{\bigg\{\frac{2\delta{\gamma}~\kappa^2\sqrt{\lambda C_1} - 3}{9\theta\sqrt{C_1}}\bigg\}}~~,
 \label{26}
\end{eqnarray}
Let us find the difference between $t_s$ and $t_\mathrm{max}$,
\begin{eqnarray}
 t_s - t_\mathrm{max} = -\frac{2}{\tilde{\eta}}~\ln{\bigg\{1 - \frac{2}{3}\delta{\gamma}~\kappa^2\sqrt{\lambda C_1}\bigg\}} > 0~~.
 \label{27}
\end{eqnarray}
Hence it follows that $t_s$ is larger than $t_\mathrm{max}$. Thus in the limit $t \rightarrow t_\mathrm{max}$, the effective energy density
and the effective pressure are given by
\begin{eqnarray}
 \rho_\mathrm{max} = \rho_\mathrm{eff}(t_\mathrm{max}) = \frac{1}{\theta^2}\bigg(\delta {\gamma} - \frac{1}{\kappa^2\sqrt{\lambda C_1}}\bigg)^2~~,
 \label{28}
\end{eqnarray}
and
\begin{eqnarray}
 \big|p_\mathrm{max}\big| = \big|p(t_\mathrm{max})\big| = \big|\omega_0\rho_\mathrm{max} - 9\tau H_\mathrm{max}^2\big|
 = \big|\omega_0 - 3\tau\kappa^2\big|\rho_\mathrm{max}~~,
 \label{29}
\end{eqnarray}
respectively. In the limit when the radiation parameter $\lambda \rightarrow \lambda_0$ where $\lambda_0$ is given by
\begin{eqnarray}
 \lambda_0 = \bigg(\frac{3}{\sqrt{\rho_0}\,\delta\,\tilde{\gamma}~\kappa^2}\bigg)^2~~,
 \label{30}
\end{eqnarray}
$\rho_\mathrm{max} \rightarrow 0$ and $\big|p_\mathrm{max}\big| \rightarrow 0$. However, the higher derivatives of the Hubble function do not
diverge. Thus, the formation of a singularity of IV type does not occur in this model.

We see that viscosity softens the singularity or avoids it altogether. This is due to a decrease in the pressure of a viscous fluid due to the viscosity term in the equation of state.

\subsection{Inhomogeneous fluid with variable parameter $\omega$}

In this section we will assume that the thermodynamic parameter $\omega(\rho,t)$ is
 a function of the energy density of the fluid. Let us choose it to have  the  form
\begin{eqnarray}
 \omega(\rho,t) = A_0\rho^{\alpha - 1} - 1~~,
 \label{31}
\end{eqnarray}
where $A_0 \neq 0$ is a dimensional constant. The   bulk viscosity is taken to be proportional to $H^n$,
\begin{eqnarray}
 \xi(H,t) = \tau\big(3H\big)^{n}
 \label{32}
\end{eqnarray}
with $\tau$ and  $n$  positive. In the case $n = 2\alpha - 1$, the energy density becomes \cite{Elizalde:2014ova}
\begin{eqnarray}
 H(t) = H_0 \bigg\{ \rho_0 \exp{[(\alpha -\frac{1}{2})\eta t]}  + \frac{\mu}{\tilde{\eta}}   \bigg\}^{\frac{2}{1-2\alpha}}~~, \quad H_0= \frac{\kappa}{\sqrt 3}\sqrt{\rho_0},
 \label{33}
\end{eqnarray}
where $\mu$ is a dimensionless constant and $\alpha \neq \frac{1}{2}$. Here $t_0$ is the present time, and $\rho_0=\rho(t_0)$, The dimension is $[\rho_0]=~$cm$^{-4}$.

 For $\alpha > \frac{1}{2}$, then at $t \rightarrow t_s
= \frac{1}{\tilde{\eta}\big(\alpha - 1/2\big)}\ln{\bigg(-\frac{\mu}{C_1\tilde{\eta}}\bigg)}$,
the Hubble parameter diverges and we obtain again the Big Rip singularity. Further, if  we consider the case $\alpha = 3/2$  the scale factor
turns out to be
\begin{eqnarray}
 a(t) = a_0\bigg(1 + \frac{\mu}{C_1 \tilde{\eta}}e^{-\eta t}\bigg)^{\frac{rH_0}{\mu}}
 \label{34}
\end{eqnarray}
and consequently the energy density in terms of the scale factor becomes
\begin{eqnarray}
 \rho_\mathrm{eff} =  \rho_0(C_1 e^{\eta t})^{-2}  \bigg[\frac{a(t)}{a_0}\bigg]^{ \frac{2\mu}{rH_0}}
 \label{35}
\end{eqnarray}
The inequality in Eq.(\ref{12}) implies the following restriction on the scale factor,
\begin{eqnarray}
 a(t) \leq a_0\bigg(\frac{ 2\kappa^2\sqrt{\lambda \rho_0}}{3C_1}\bigg)
 ^{\frac{rH_0}{\mu} }
 \exp[\frac{ rH_0\eta t}{\mu}]
 \label{36}
\end{eqnarray}
Scale factor values are limited by the maximum number $a_\mathrm{max}$,
\begin{eqnarray}
 a(t) \leq a_\mathrm{max} = a_0\bigg(\frac{2\kappa^2\sqrt{\lambda \rho_0}}{3C_1}\bigg)^{\frac{rH_0}{\mu}}
 \label{37}
\end{eqnarray}
which corresponds to the  time
\begin{eqnarray}
 t_\mathrm{max} = \frac{1}{\eta}~\ln\frac{3\mu}{\tilde{\eta}\big(2\kappa^2 \sqrt{\lambda \rho_0} - 3C_1)} ~~.
 \label{38}
\end{eqnarray}
The difference between $t_s$ and $t_\mathrm{max}$ takes the following form,
\begin{eqnarray}
 t_s - t_\mathrm{max} = \frac{1}{\eta}~\ln\big(1 - \frac{2\kappa^2}{3C_1}\sqrt{\lambda\rho_0}\big) < 0~~.
 \label{39}
\end{eqnarray}
It turns out that $t_\mathrm{max}$ is larger than $t_s$. Thus, it  due to thermal radiation, the
time of formation of the cosmological singularity changes qualitatively. In the limit $t \rightarrow t_\mathrm{max}$, the effective energy density and
the effective pressure are given  by
\begin{eqnarray}
 \rho_\mathrm{max} = \rho_\mathrm{eff}(t_\mathrm{max}) =\rho_0\left[ \frac{\tilde{\eta}}{\mu}\left( 1-\frac{3C_1}{2\kappa^2\sqrt{\lambda \rho_0}}\right) \right],
 \label{40}
\end{eqnarray}
and
\begin{eqnarray}
 \big|p_\mathrm{max}\big| = \big|p(t_\mathrm{max})\big| = \bigg|\bigg(A_0\rho_\mathrm{max}^{\frac{1}{2}} - 1\bigg)\rho_\mathrm{max}
 - \tau\big(3H_\mathrm{max}\big)^3\bigg| = \bigg|\bigg(A_0 - 3\sqrt{3}~\tau\kappa^3\bigg)\rho_\mathrm{max}^{\frac{1}{2}} - 1\bigg|\rho_\mathrm{max}~~,
 \label{41}
\end{eqnarray}
respectively. Thereby, the value of the energy density and effective pressure becomes finite, while higher derivatives
of the Hubble function diverge. Consequently, a cosmological singularity is not formed.

\section{Conclusion}

 We have investigated the singular behavior of the dark universe, taking into
account the thermal effects caused by Hawking radiation, the viscosity properties of the dark
fluid, and its interaction with dark matter on the visible horizon of the FRW universe. According
to the study carried out in \cite{Nojiri:2020sti} for an ideal fluid, near the singularity it is necessary to take into
account the Hawking thermal radiation, which leads to a change in the type of the singularity. In
a dark universe with singularities of types I and III with a finite formation time, a transition to a type II singularity occurs.

Models with an inhomogeneous viscous dark fluid interacting with dark matter were considered
in \cite{Elizalde:2014ova}. It is shown that, in our case, a transition from singularity of type I to singularity III
type is possible due to the influence of thermal radiation. Singularities may be absent, due
 to the viscosity of the dark fluid and its interaction with
dark matter. The absence of a singularity in some models is explained by the fact that both the
presence of a bulk  viscosity in the equation of state of a dark fluid, and the presence of a thermal radiation term
in the Friedmann equation, are proportional to a power of the Hubble parameter. As a result,
the effect of thermal radiation near the singularity may be  neutralized by a viscous fluid.

As was shown in Ref.~\cite{Astashenok:2021}, taking thermal radiation into account does not weaken the agreement of cosmological models with astronomical obeservations.

%%%%%%%%%%%%%%%%%%%%%%%%%%%%%%%%%%%%%%%%%%%%%%%%%%%%%%%%%%%%%%%%%%%%%%%%%%%%
\subsection*{Acknowledgments}
This work was  supported in part by Ministry of Education of Russian Federation,  Project No FEWF-2020-0003 (A.V.T.).
%%%%%%%%%%%%%%%%%%%%%%%%%%%%%%%%%%%%%%%%%%%%%%%%%%%%%%%%%%%%%%%%%%%%%%%%%%%%


\begin{thebibliography}{99}



 %\cite{Nojiri:2017ncd}
\bibitem{Nojiri:2017ncd}
S.~Nojiri, S.~D.~Odintsov and V.~K.~Oikonomou,
Modified gravity theories on a nutshell: Inflation, bounce and late-time evolution,
Phys. Rept. \textbf{692} (2017), 1-104
doi:10.1016/j.physrep.2017.06.001
[arXiv:1705.11098 [gr-qc]].
%832 citations counted in INSPIRE as of 30 Dec 2020



%\cite{Shi:2011sa}
\bibitem{Shi:2011sa}
Y.~Shi,
A cyclic cosmological model based on the   $f(\rho)$     modified theory of gravity,
[arXiv:1106.0341 [physics.gen-ph]].
%0 citations counted in INSPIRE as of 30 Dec 2020


%\cite{Perivolaropoulos:2006ce}
\bibitem{Perivolaropoulos:2006ce}
L.~Perivolaropoulos,
Accelerating universe: observational status and theoretical implications,
AIP Conf. Proc. \textbf{848} (2006) no.1, 698-712
doi:10.1063/1.2348048
[arXiv:astro-ph/0601014 [astro-ph]].
%140 citations counted in INSPIRE as of 30 Dec 2020



%\cite{Arefeva:2006ido}
\bibitem{Arefeva:2006ido}
I.~Y.~Aref'eva and I.~V.~Volovich,
On the null energy condition and cosmology,
Theor. Math. Phys. \textbf{155} (2008), 503-511
doi:10.1007/s11232-008-0041-8
[arXiv:hep-th/0612098 [hep-th]].
%77 citations counted in INSPIRE as of 30 Dec 2020


%\cite{Nojiri:2005sx}
\bibitem{Nojiri:2005sx}
S.~Nojiri, S.~D.~Odintsov and S.~Tsujikawa,
Properties of singularities in (phantom) dark energy universe,
Phys. Rev. D \textbf{71} (2005), 063004
doi:10.1103/PhysRevD.71.063004
[arXiv:hep-th/0501025 [hep-th]].
%1003 citations counted in INSPIRE as of 30 Dec 2020


%\cite{Nojiri:2006zh}
\bibitem{Nojiri:2006zh}
S.~Nojiri and S.~D.~Odintsov,
The new form of the equation of state for dark energy fluid and accelerating universe,
Phys. Lett. B \textbf{639} (2006), 144-150
doi:10.1016/j.physletb.2006.06.065
[arXiv:hep-th/0606025 [hep-th]].
%157 citations counted in INSPIRE as of 30 Dec 2020


%\cite{Nojiri:2004pf}
\bibitem{Nojiri:2004pf}
S.~Nojiri and S.~D.~Odintsov,
The final state and thermodynamics of dark energy universe,
Phys. Rev. D \textbf{70} (2004), 103522
doi:10.1103/PhysRevD.70.103522
[arXiv:hep-th/0408170 [hep-th]].
%505 citations counted in INSPIRE as of 30 Dec 2020


%\cite{Setare:2007bx}
\bibitem{Setare:2007bx}
M.~R.~Setare,
Interacting generalized Chaplygin gas model in non-flat universe,
Eur. Phys. J. C \textbf{52} (2007), 689-692
doi:10.1140/epjc/s10052-007-0405-5
[arXiv:0711.0524 [gr-qc]].
%115 citations counted in INSPIRE as of 30 Dec 2020


%\cite{Capozziello:2005mj}
\bibitem{Capozziello:2005mj}
S.~Capozziello, S.~Nojiri and S.~D.~Odintsov,
Dark energy: The equation of state description versus scalar-tensor or modified gravity,
Phys. Lett. B \textbf{634} (2006), 93-100
doi:10.1016/j.physletb.2006.01.065
[arXiv:hep-th/0512118 [hep-th]].
%208 citations counted in INSPIRE as of 30 Dec 2020


%\cite{Cognola:2006eg}
\bibitem{Cognola:2006eg}
G.~Cognola, E.~Elizalde, S.~Nojiri, S.~D.~Odintsov and S.~Zerbini,
Dark energy in modified Gauss-Bonnet gravity: Late-time acceleration and the hierarchy problem,
Phys. Rev. D \textbf{73} (2006), 084007
doi:10.1103/PhysRevD.73.084007
[arXiv:hep-th/0601008 [hep-th]].
%547 citations counted in INSPIRE as of 30 Dec 2020


%\cite{Nojiri:2005sr}
\bibitem{Nojiri:2005sr}
S.~Nojiri and S.~D.~Odintsov,
Inhomogeneous equation of state of the universe: Phantom era, future singularity and crossing the phantom barrier,
Phys. Rev. D \textbf{72} (2005), 023003
doi:10.1103/PhysRevD.72.023003
[arXiv:hep-th/0505215 [hep-th]].
%614 citations counted in INSPIRE as of 30 Dec 2020

%\cite{Nojiri:2020wmh}
\bibitem{Nojiri:2020wmh}
S.~Nojiri, S.~D.~Odintsov, V.~K.~Oikonomou and T.~Paul,
Unifying holographic inflation with holographic dark energy: a covariant approach,
Phys. Rev. D \textbf{102} (2020) no.2, 023540
doi:10.1103/PhysRevD.102.023540
[arXiv:2007.06829 [gr-qc]].
%4 citations counted in INSPIRE as of 30 Dec 2020


%\cite{Odintsov:2020zct}
\bibitem{Odintsov:2020zct}
S.~D.~Odintsov, V.~K.~Oikonomou and T.~Paul,
From a bounce to the dark energy era with $F(R)$ gravity,
Class. Quant. Grav. \textbf{37} (2020) no.23, 235005
doi:10.1088/1361-6382/abbc47
[arXiv:2009.09947 [gr-qc]].
%2 citations counted in INSPIRE as of 30 Dec 2020



%\cite{Nakamura:2010zzi}
\bibitem{Nakamura:2010zzi}
K.~Nakamura \textit{et al.} [Particle Data Group],
Review of particle physics,
J. Phys. G \textbf{37} (2010), 075021
doi:10.1088/0954-3899/37/7A/075021
%6559 citations counted in INSPIRE as of 30 Dec 2020


%\cite{Caldwell:1999ew}
\bibitem{Caldwell:1999ew}
R.~R.~Caldwell,
A phantom menace?,
Phys. Lett. B \textbf{545} (2002), 23-29
doi:10.1016/S0370-2693(02)02589-3
[arXiv:astro-ph/9908168 [astro-ph]].
%2599 citations counted in INSPIRE as of 30 Dec 2020


%\cite{Cai:2009ph}
\bibitem{Cai:2009ph}
R.~G.~Cai and N.~Ohta,
Horizon thermodynamics and gravitational field equations in Horava-Lifshitz Gravity,
Phys. Rev. D \textbf{81} (2010), 084061
doi:10.1103/PhysRevD.81.084061
[arXiv:0910.2307 [hep-th]].
%122 citations counted in INSPIRE as of 30 Dec 2020


%\cite{Nojiri:2020sti}
\bibitem{Nojiri:2020sti}
S.~Nojiri and S.~D.~Odintsov,
The dark universe future and singularities: the account of thermal and quantum effects,
Phys. Dark Univ. \textbf{30} (2020), 100695
doi:10.1016/j.dark.2020.100695
[arXiv:2006.03946 [gr-qc]]; A.~Astashenok, S.~D.~Odintsov and V.~K.~Oikonomou, submitted.
%2 citations counted in INSPIRE as of 30 Dec 2020



%\cite{Gron:1990ew}
\bibitem{Gron:1990ew}
{\O}.~Gr{\o}n,
Viscous inflationary universe models,
Astrophys. Space Sci. \textbf{173} (1990), 191-225
doi:10.1007/BF00643930
%159 citations counted in INSPIRE as of 30 Dec 2020


%\cite{Brevik:2017msy}
\bibitem{Brevik:2017msy}
I.~Brevik, \O{}.~Gr\o{}n, J.~de Haro, S.~D.~Odintsov and E.~N.~Saridakis,
Viscous cosmology for early- and late-time Universe,
Int. J. Mod. Phys. D \textbf{26} (2017) no.14, 1730024
doi:10.1142/S0218271817300245
[arXiv:1706.02543 [gr-qc]].
%100 citations counted in INSPIRE as of 30 Dec 2020


%\cite{Cataldo:2005qh}
\bibitem{Cataldo:2005qh}
M.~Cataldo, N.~Cruz and S.~Lepe,
Viscous dark energy and phantom evolution,
Phys. Lett. B \textbf{619} (2005), 5-10
doi:10.1016/j.physletb.2005.05.029
[arXiv:hep-th/0506153 [hep-th]].
%100 citations counted in INSPIRE as of 30 Dec 2020


%\cite{Brevik:2005gx}
\bibitem{Brevik:2005gx}
I.~Brevik, J.~M.~B{\o}rven and S.~Ng,
Viscous brane cosmology with a brane-bulk energy interchange term,
Gen. Rel. Grav. \textbf{38} (2006), 907-915
doi:10.1007/s10714-006-0271-8
[arXiv:gr-qc/0512026 [gr-qc]].
%25 citations counted in INSPIRE as of 30 Dec 2020


%\cite{Brevik:2001ed}
\bibitem{Brevik:2001ed}
I.~Brevik and S.~D.~Odintsov,
On the Cardy-Verlinde entropy formula in viscous cosmology,
Phys. Rev. D \textbf{65} (2002), 067302
doi:10.1103/PhysRevD.65.067302
[arXiv:gr-qc/0110105 [gr-qc]].
%67 citations counted in INSPIRE as of 30 Dec 2020


%\cite{Brevik:2010jv}
\bibitem{Brevik:2010jv}
I.~Brevik, S.~Nojiri, S.~D.~Odintsov and D.~Saez-Gomez,
Cardy-Verlinde formula in FRW Universe with inhomogeneous generalized fluid and dynamical entropy bounds near the future singularity,
Eur. Phys. J. C \textbf{69} (2010), 563-574
doi:10.1140/epjc/s10052-010-1425-0
[arXiv:1002.1942 [hep-th]].
%38 citations counted in INSPIRE as of 30 Dec 2020


%\cite{Brevik:2005bj}
\bibitem{Brevik:2005bj}
I.~Brevik and O.~Gorbunova,
Dark energy and viscous cosmology,
Gen. Rel. Grav. \textbf{37} (2005), 2039-2045
doi:10.1007/s10714-005-0178-9
[arXiv:gr-qc/0504001 [gr-qc]].
%205 citations counted in INSPIRE as of 30 Dec 2020


%\cite{Li:2009mf}
\bibitem{Li:2009mf}
B.~Li and J.~D.~Barrow,
Does bulk viscosity create a viable unified dark matter model?,
Phys. Rev. D \textbf{79} (2009), 103521
doi:10.1103/PhysRevD.79.103521
[arXiv:0902.3163 [gr-qc]].
%132 citations counted in INSPIRE as of 30 Dec 2020


%\cite{Brevik:2017juz}
\bibitem{Brevik:2017juz}
I.~Brevik, E.~Elizalde, S.~D.~Odintsov and A.~V.~Timoshkin,
Inflationary universe in terms of a van der Waals viscous fluid,
Int. J. Geom. Meth. Mod. Phys. \textbf{14} (2017) no.12, 1750185
doi:10.1142/S0219887817501857
[arXiv:1708.06244 [gr-qc]].
%30 citations counted in INSPIRE as of 30 Dec 2020


%\cite{Nojiri:2003vn}
\bibitem{Nojiri:2003vn}
S.~Nojiri and S.~D.~Odintsov,
Quantum de Sitter cosmology and phantom matter,
Phys. Lett. B \textbf{562} (2003), 147-152
doi:10.1016/S0370-2693(03)00594-X
[arXiv:hep-th/0303117 [hep-th]].
%776 citations counted in INSPIRE as of 30 Dec 2020


%\cite{Barrow:2013ria}
\bibitem{Barrow:2013ria}
J.~D.~Barrow and S.~Cotsakis,
Geodesics at sudden singularities,
Phys. Rev. D \textbf{88} (2013), 067301
doi:10.1103/PhysRevD.88.067301
[arXiv:1307.5005 [gr-qc]].
%25 citations counted in INSPIRE as of 30 Dec 2020


%\cite{Nojiri:2004ip}
\bibitem{Nojiri:2004ip}
S.~Nojiri and S.~D.~Odintsov,
Quantum escape of sudden future singularity,
Phys. Lett. B \textbf{595} (2004), 1-8
doi:10.1016/j.physletb.2004.06.060
[arXiv:hep-th/0405078 [hep-th]].
%301 citations counted in INSPIRE as of 30 Dec 2020


%\cite{Myrzakul:2013qka}
\bibitem{Myrzakul:2013qka}
S.~Myrzakul, R.~Myrzakulov and L.~Sebastiani,
Inhomogeneous viscous fluids in FRW universe and finite-future time singularities,
Astrophys. Space Sci. \textbf{350} (2014), 845-853
doi:10.1007/s10509-014-1799-9
[arXiv:1311.6939 [gr-qc]].
%21 citations counted in INSPIRE as of 30 Dec 2020


%\cite{Capozziello:2009hc}
\bibitem{Capozziello:2009hc}
S.~Capozziello, M.~De Laurentis, S.~Nojiri and S.~D.~Odintsov,
Classifying and avoiding singularities in the alternative gravity dark energy models,
Phys. Rev. D \textbf{79} (2009), 124007
doi:10.1103/PhysRevD.79.124007
[arXiv:0903.2753 [hep-th]].
%107 citations counted in INSPIRE as of 30 Dec 2020


%\cite{Elizalde:2014ova}
\bibitem{Elizalde:2014ova}
E.~Elizalde, V.~V.~Obukhov and A.~V.~Timoshkin,
Inhomogeneous viscous dark fluid coupled with dark matter in the FRW universe,
Mod. Phys. Lett. A \textbf{29} (2014) no.25, 1450132
doi:10.1142/S0217732314501326
[arXiv:1406.7653 [gr-qc]].
%20 citations counted in INSPIRE as of 30 Dec 2020


%\cite{Odintsov:2018uaw}
\bibitem{Odintsov:2018uaw}
S.~D.~Odintsov and V.~K.~Oikonomou,
Dynamical Systems Perspective of Cosmological Finite-time Singularities in $f(R)$ Gravity and Interacting Multifluid Cosmology,
Phys. Rev. D \textbf{98} (2018) no.2, 024013
doi:10.1103/PhysRevD.98.024013
[arXiv:1806.07295 [gr-qc]].
%34 citations counted in INSPIRE as of 30 Dec 2020


%\cite{Frampton:2011aa}
\bibitem{Frampton:2011aa}
P.~H.~Frampton, K.~J.~Ludwick and R.~J.~Scherrer,
Pseudo-rip: Cosmological models intermediate between the cosmological constant and the little rip,
Phys. Rev. D \textbf{85} (2012), 083001
doi:10.1103/PhysRevD.85.083001
[arXiv:1112.2964 [astro-ph.CO]].
%95 citations counted in INSPIRE as of 30 Dec 2020


%\cite{Frampton:2011sp}
\bibitem{Frampton:2011sp}
P.~H.~Frampton, K.~J.~Ludwick and R.~J.~Scherrer,
The Little Rip,
Phys. Rev. D \textbf{84} (2011), 063003
doi:10.1103/PhysRevD.84.063003
[arXiv:1106.4996 [astro-ph.CO]].
%196 citations counted in INSPIRE as of 30 Dec 2020


%\cite{Frampton:2011rh}
\bibitem{Frampton:2011rh}
P.~H.~Frampton, K.~J.~Ludwick, S.~Nojiri, S.~D.~Odintsov and R.~J.~Scherrer,
Models for Little Rip dark energy,
Phys. Lett. B \textbf{708} (2012), 204-211
doi:10.1016/j.physletb.2012.01.048
[arXiv:1108.0067 [hep-th]].
%107 citations counted in INSPIRE as of 30 Dec 2020


%\cite{Capozziello:2005pa}
\bibitem{Capozziello:2005pa}
S.~Capozziello, V.~F.~Cardone, E.~Elizalde, S.~Nojiri and S.~D.~Odintsov,
Observational constraints on dark energy with generalized equations of state,
Phys. Rev. D \textbf{73} (2006), 043512
doi:10.1103/PhysRevD.73.043512
[arXiv:astro-ph/0508350 [astro-ph]].
%304 citations counted in INSPIRE as of 30 Dec 2020


\bibitem{Astashenok:2021}
V.~Astashenok, S.~D.~Odintsov and V.~K.~Oikonomou,
Dark energy and cosmological horizon thermal effects,
Phys. Rev. D \textbf{103} (2021), 043514
doi:10.1103/PhysRevD.103.043514.







\end{thebibliography}
\end{document}